\documentclass[9pt,twocolumn,twoside]{osajnl}

\usepackage{amsmath}

\journal{ol} 

\setboolean{shortarticle}{true} 

\title{Phase retrieval with physics informed zero-shot learning}

\author[]{Sanjeev Kumar}

\affil[]{School of Medical Science and Technology, Indian Institute of Technology, Kharagpur, 721302, India}

\affil[]{sanjeevsmst@iitkgp.ac.in}

\begin{abstract}
Phase can be reliably estimated from a single diffracted intensity image, if a faithful prior information about the object is available. Examples include amplitude bounds, object support, sparsity in the spatial or a transform domain, deep image prior and the prior learnt from the labelled datasets by a deep neural network. Deep learning facilitates state of art reconstruction quality but requires a large labelled dataset (ground truth-measurement pair acquired in the same experimental conditions) for training. To alleviate this data requirement problem, this letter proposes a zero-shot learning method. The letter demonstrates that the object-prior learnt by a deep neural network while being trained for a denoising task can also be utilized for the phase retrieval, if the diffraction physics is effectively enforced on the network output. The letter additionally demonstrates that the incorporation of total variation in the proposed zero-shot framework facilitates the reconstruction of similar quality in lesser time ( e.g. $\sim$8.5 fold, for a test reported in this letter).
\end{abstract}

\setboolean{displaycopyright}{true}

\begin{document}

\maketitle

Phase retrieval from a single diffracted intensity image is an ill-posed inverse problem which can be solved using an iterative algorithm or a pre-trained deep neural network, in coherent or partially coherent illuminations \cite{fienup1982phase, rivenson2018phase}. It enables imaging of transparent or nearly-transparent samples, commonly referred as weakly scattering samples, which otherwise do not produce sufficient contrast in the intensity images. Phase imaging is of utter importance in the live-cell monitoring and histopathological microscopy of unstained tissue samples.

As any other ill-posed inverse problem, the choice of object-prior plays an integral role in determining the quality of solution obtained in a phase retrieval problem. In the earlier iterative methods like the Gerchberg-Saxton's method \cite{gerchberg1972practical} and the Fienup's method \cite{fienup1982phase}, the amplitude constraints \cite{latychevskaia2007solution} or the object supports (the presence of features only in parts of the field of view) \cite{mudanyali2010compact} were chosen as the priors. Sparsity in the total-variation (TV) domain, or in a wavelet domain or in a domain learnt from data (dictionary learning) were established as excellent hand-crafted object-priors in the image restoration problems like degradation and denoising \cite{chambolle2004algorithm,bioucas2007new}. 
Katkovnik et al, 2011 used sparse modelling of both the amplitude and the phase for complex object estimation \cite{katkovnik2012high}. Some other hand-crafted priors were used by Jolivet et al, 2018 in their optimization based reconstruction framework, for in-line holography \cite{jolivet2018regularized}. They included bounds on the transmittance values, maximum and minimum phase and spatial smoothness.

Rivenson et al, 2018 reported phase imaging in holography through end-to-end learning with a deep convolutional neural network (deep CNN) \cite{rivenson2018phase,rivenson2019deep,wu2018extended}. This network was trained with ground truth images obtained by multi-height phase recovery method \cite{misell1973examination, paganin2004quantitative} (8 multi-height holograms per image were used). Wang et al, 2019 reported another deep network for the same task but they trained their network with ground truth phase images reconstructed by the convolution method \cite{wang2019net}. In 2018, Ulyanov et al reported the use of untrained deep CNN as a handcrafted prior and called it "deep image prior" (DIP) \cite{ulyanov2018deep}. They demonstrated that the natural structure of the deep CNN promotes the natural signal and avoids the noise. While training, it inherently poses a higher impedence to the values of network parameters which will output the noise than the natural images. They demonstrated that network will represent noise only if it is extensively trained to do so. In 2020, Wang et al demonstrated that the deep image prior can also be used for reconstructing the phase images \cite{wang2020phase}. They demonstrated that a deep CNN similarly poses a higher impedence to the values of network parameters which will output the artifacts associated with the missing phase information.

End-to-end learning frameworks have the burden of learning both the physics i.e. the light transport and measurement model as well as the prior. Alternatively, Physics informed (or physics aware) learning frameworks integrate the partially or fully available physical or mathematical models with the data-driven regression methods \cite{barbastathis2019use, karniadakis2021physics}. The additional knowledge from the former models can act as a regularizer to constrain the space of admissible solutions to a smaller and manageable size \cite{raissi2019physics}. The motivation is to improve the generalization \cite{heaven2019deep}, reduce the big data requirement or training time and achieve a better accuracy \cite{karniadakis2021physics, raissi2019physics}. In 1992, Psichogios et al, proposed a hybrid neural network where a neural network estimates the unknown parameters of interest, which are fed to a model predefined using the known physical considerations \cite{psichogios1992hybrid}. Raissi et al, 2019 similarly applied a neural network and their mathematical model in series, specifically in this order to constraint the final output in a smaller admissible solution space \cite{raissi2019physics}. Chan et al, 2021 used this principle in coherent diffractive imaging. They used their physics-aware optimization procedure as a refinement step on the deep network's output. They trained their network on simulated data, obtained using a physics-informed data preparation pipeline to serve as a representative of experimental data \cite{chan2021rapid}.


\begin{figure}
\centering
{\includegraphics[width=8cm]{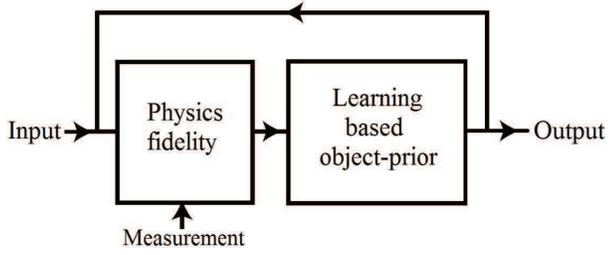}}
\caption{Recurrent approach of physics informed deep learning for phase retrieval. Each iteration enforces the 
 known physical laws and the deep prior learnt from the data.}
\label{Fig:1}
\end{figure}

Unlike the previous physics informed machine learning methods, this letter combines the knowledge of physical model and deep learning in a recurrent manner. The letter uses a deep neural network originally trained for the removal of additive white Gaussian noise (AWGN) to retrieve the phase information from a diffracted intensity recording, without any fine-tuning of network parameters. This transfer of knowledge has been achieved here by enforcing the additional knowledge of diffraction physics in a recurrent manner. The letter first presents a method labelled Phy-ZSN (short for physics informed zero-shot network), which combines the classical error-reduction algorithm and deep learning for obtaining more accurate phase estimations. The method outperforms the phase retrieval with total-variation (TV) prior in terms of image quality but at the cost of larger reconstruction time. To alleviate this problem, this letter presents a second method labelled PhyTV-ZSN, which combines both the handcrafted prior TV and the deep neural network to obtain the improved reconstruction accuracy in a significantly less time. Unlike other deep learning based phase reconstruction methods, the proposed method does not need any labelled phase image data (ground truth). 

The forward diffraction model can be represented as:
\begin{equation}
I = \wp(|g|^2), \ \textnormal{where} \ g = f \otimes h(z,\lambda)
\end{equation}
$I$ is the recorded intensity image (diffraction pattern), $\wp$ is an operator accounting for the noise of unknown statistics, $g$ is the complex wave-field at the sensor plane, $f$ is the complex wave-field at the object plane and $h$ is the depth and wavelength dependent impulse response function. $z$ denotes the distance from the object to the sensor plane and $\lambda$ denotes the wavelength of the light. $|.|$ denotes the amplitude part of the complex quantity (.) and $\otimes$ denotes the convolution operation. Under the assumption of spatial and temporal coherence, impulse response function $h(z,\lambda)$ is calculated using the scalar-diffraction equations \cite{goodman2005introduction}. The problem statement is to estimate the complex object $f$ from the measurement $I$.  Additionally, the single scattering (first Born-approximation \cite{born2013principles}) and single depth approximations have been followed in this letter.

Zero-shot learning (ZSL) belongs to the family of transfer learning \cite{wang2019survey} where a model performs the given task on samples from classes which were never observed during the training phase. In ZSL, to facilitate this broader generalization, some additional information about these unseen classes is provided in the form of auxiliary information \cite{xian2018zero}. For example, a neural network trained to identify horses can be used to identify zebras if the auxiliary information is provided that zebras look like striped horses \cite{wang2019survey}. The proposed method uses a deep CNN trained for the additive white Gaussian noise (AWGN) removal \cite{zhang2020plug, zhang2017learning}, and transfers this
learning to the phase retrieval problem. The proposed framework operates in the recurrent manner shown in the figure 1. The Physics fidelity block enforces the known diffraction physics and measurement model (the auxiliary information in ZSL), followed by passing the estimate through the pre-trained deep CNN for enforcing the learnt object-prior. The process is repeated until the convergence curve approaches a stationary region. The two blocks will be discussed in details below.

\begin{figure}
\centering
{\includegraphics[width=8cm]{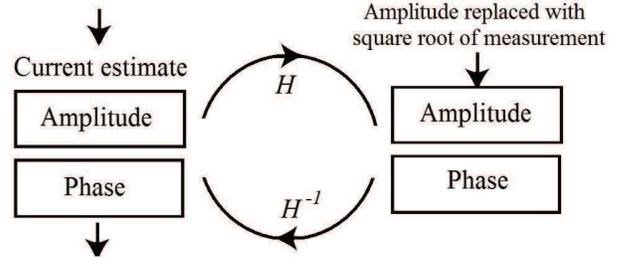}}
\caption{Physics fidelity block based on the error-reduction algorithm.}
\label{Fig:2}
\end{figure}
\begin{figure}
\centering
{\includegraphics[width=8.5cm]{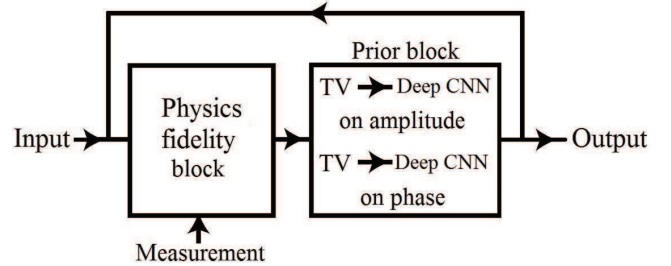}}
\caption{PhyTV-ZSN: The object-prior block shows the cascaded TV and ZSN priors. PhyTV-ZSN is identical to the Phy-ZSN except that the two TV units are absent in the latter.}
\label{Fig:3}
\end{figure}
\begin{figure*}[t]
\centering
{\includegraphics[width=17.5cm]{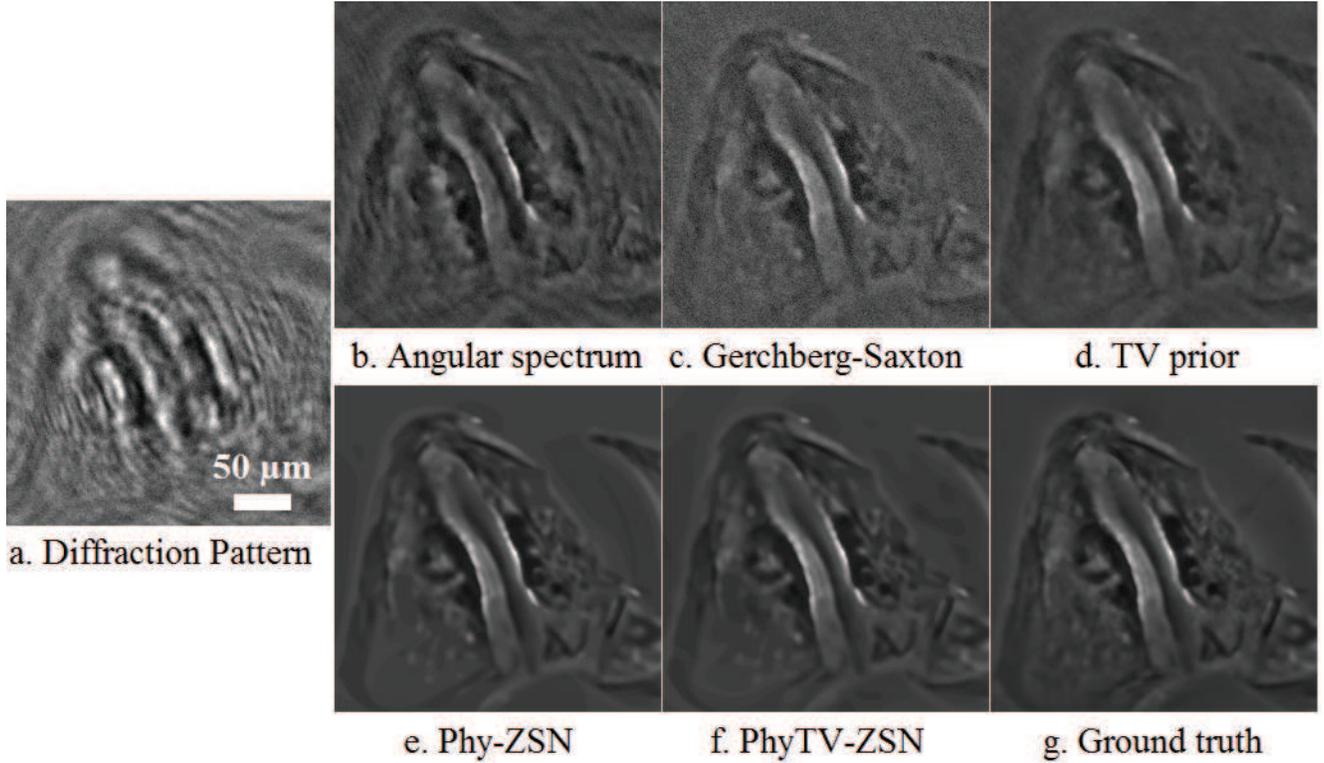}}
\caption{ (a) The Poisson noise corrupted diffraction pattern of the object "Cell" at $z$ = 1 mm. The reconstructed phase images with (b) Angular spectrum wave propagation method. (c) Gerchberg-Saxton method (amplitude constriant). (d) Error-reduction method with TV prior applied to both amplitude and phase. (e) Phy-ZSN (proposed method 1). (f) PhyTV-ZSN (proposed method 2). (g) Ground truth. Phase values were in the range 0 to 2$\pi$ but have been scaled to the range 0 to 255 for visualization.}
\label{Fig:4}
\end{figure*}
\begin{figure*}[t]
\centering
{\includegraphics[width=17.5cm]{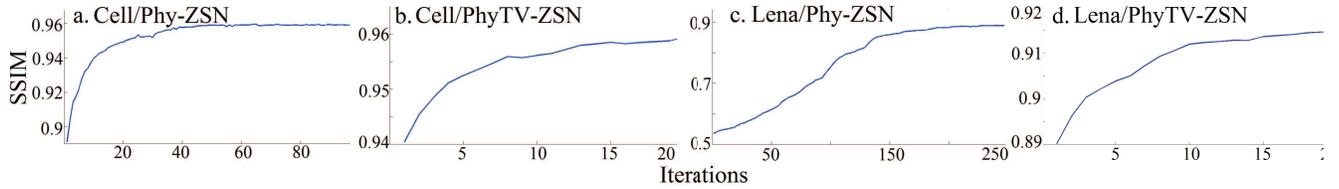}}
\caption{Convergence pattern for the reconstructions: (a) Cell at $z$ = 1 mm, reconstructed with Phy-ZSN and (b) with PhyTV-ZSN. (c) Lena at $z$ = 0.43 mm, reconstructed with Phy-ZSN and (d) with PhyTV-ZSN. For PhyTV-ZSN (b and d), the initial estimate is the reconstruction with TV prior.}
\label{Fig:5}
\end{figure*}

In physics fidelity block, the error-reduction algorithm has been adapted. The algorithm has been shown equivalent to the steepest-descent gradient search method in the literature \cite{fienup1982phase}. Rigorous mathematical analysis on the convergence behaviour of this algorithm can be found in the reference \cite{bauschke2002phase}. The steps have been illustrated in figure 2. The object estimate is projected to the sensor plane using the impulse response function $h(z,\lambda)$ via the point-wise multiplication in the Fourier domain, amplitude is replaced by the square-root of the measurement and finally backprojected to the object plane via the point-wise division in the Fourier domain. Unlike most of impulse response functions encountered in imaging, the $h(z,\lambda)$ in the frequency domain $H(z,\lambda)$ is a phase only function. And hence the backprojection using division operation in Fourier domain is possible (i.e. well-conditioned, noise is not amplified by this division). $H(z,\lambda)$ is commonly referred as the optical transfer function (OTF). The magnitude of $H(z,\lambda)$ is one in the supported bandwidth and zero outside. Hence, the sampling rate must be capped by this limit. 

For free space (refractive index, $n=1$), the optical transfer function is obtained by the following equation \cite{ozcan2016lensless}:
\begin{equation}
H(\textbf{v}) =  exp\Big(jk_0z \sqrt{1 - (\lambda v_x)^2-(\lambda v_y)^2}\Big); \ v^2_x + v^2_y < \frac{1}{\lambda^2} 
\end{equation}
\begin{equation}
H(\textbf{v}) =  0;\ \textnormal{otherwise} 
\end{equation}
where, $\textbf{v}=(v_x,v_y)$ is the frequency coordinate vector, $k_0 = \frac{2 \pi}{\lambda}$ is the wave-number.

\begin{table}[t]
\centering
\caption{\bf Quantitative assessment of reconstructions.}
    \resizebox{8cm}{!}{%
    \begin{tabular}{|l|l|l|l|l|}
  \hline      
  \rule{0pt}{1\normalbaselineskip} 
   & \multicolumn{2}{c|}{\bf{Cell}}  & \multicolumn{2}{c|}{\bf{Lena}} \\[2pt]
  \hline
  \rule{0pt}{1\normalbaselineskip} 
  \bf Method & \bf SSIM & \bf PSNR & \bf SSIM & \bf PSNR\\[2pt]
  \hline
  \rule{0pt}{1\normalbaselineskip} 
  Angular spectrum Prop. & 0.6692  & 25.19 & 0.4740 & 13.59 \\
  \rule{0pt}{1\normalbaselineskip}
  Gerchberg-Saxton method & 0.4438 & 20.21 & 0.2990 & 15.00  \\
  \rule{0pt}{1\normalbaselineskip}
  Error-reduction with TV  & 0.9161 & 29.24 & 0.8630 & 25.41\\
  \rule{0pt}{1\normalbaselineskip} 
  Phy-ZSN  & 0.9629 & 34.45 & 0.8875 & 23.16\\
  \rule{0pt}{1\normalbaselineskip} 
  PhyTV-ZSN  & 0.9601 & 34.84 & 0.9145 & 27.32 \\[1pt]
  \hline
  \end{tabular}
}
\end{table}

In learning based object-prior block in figure 1, a deep CNN based on the U-Net and the ResNet architecture, referred as the DRUNet \cite{zhang2020plug,zhang2017learning} has been used. The network was trained for denoising AWGN of randomly chosen variances in the range [0,50]. The detailed steps for training this neural network can be found in the reference \cite{zhang2020plug}. This letter uses the terminology zero-shot learning/network (ZSL/ZSN) because: (1) This trained network has not seen any image from the class considered in this letter i.e. the estimates in a phase retrieval process (figure 4b for instance is a backprojected diffraction pattern without any object prior). (2) All the simulated diffraction patterns used in this letter were corrupted with Poisson noise (not with AWGN, which was used during the training).

The amplitude and phase values obtained from the physics fidelity block are fed separately to this deep CNN. The inputs must be scaled in the range between 0 to 255 because the network was trained with the examples of intensity values in this specific range. The output must be scaled back to the original range, by the applying the inverse of previous scaling.

 As compared to the handcrafted prior TV, the deep neural network has a larger computational complexity. To compare the reconstruction time, 10 iterations of error reduction algorithm with TV prior take $\sim$1.3 seconds while the same number of iterations of Phy-ZSN take $\sim$3.5 minutes. The testing was performed on a system with 2.20 GHz Intel(R) Xeon(R) CPU and 13 GB RAM. To alleviate this problem and for additional hand-crafted prior based regularization control, another method which combines the two priors (TV and deep network) in a cascaded manner has been proposed in this letter. This method has been labelled as PhyTV-ZSN and has been shown in figure 3.

For faster reconstruction than Phy-ZSN, the following two-step implementation with PhyTV-ZSN can be used:

\textbf{\textit{Step 1:}} Disable the deep CNN part in PhyTV-ZSN until the iterative method converges to a stationary point. This is a solution with TV prior.

\textbf{\textit{Step 2:}} Enable the CNN part in the same. The iterative method will converge to another stationary point. This is a solution with TV and deep prior.

10 iterations of PhyTV-ZSN take $\sim$3.67 minutes.

Figure 4 shows the phase reconstructions with different phase retrieval methods, for comparison. The object is a HeLa cell phase image captured by an Olympus IX83 microscope using a 20x/0.4 Ph2 Objective \cite{burri2019} and will be referred as the "Cell" in the rest of this letter. The diffraction pattern was simulated by the forward diffraction model described earlier, with the following parameter values: pixel-pitch = 1.12 $\mu m$, light wavelength = 670 nm and object to sensor distance, $z$ = 1 mm. The simulated diffraction pattern was corrupted by the Poisson noise. For figure 4d, TV prior implementation was obtained using Chambolle's method \cite{chambolle2004algorithm} and the TV prior was applied on both the amplitude and phase separately. As the figure 4 and table 1 show, the proposed methods Phy-ZSN and PhyTV-ZSN give improved reconstruction accuracies (TV is considered one of the best hand-crafted object prior in the image processing community). The SSIM (structural similarity index) and PSNR (peak signal-to-noise ratio) values are very low for Gerchberg-Saxton method because of the non-suppression of the noise in the reconstructed images. The convergence patterns have been shown in figure 5. In figure 5 and table 1, the results corresponding to an another simulated diffraction pattern from an object "Lena" at $z$ = 0.43 mm have also been included. For Lena, when PhyTV-ZSN method was used, the total reconstruction time reduced by a factor of 8.5, as compared to the Phy-ZSN. The convergence rates are different for Lena and Cell because of the two different $z$ values, i.e. because of the distance dependence of the impulse response function. The Fienup's hybrid input-output algorithm can also be used in the Physics fidelity block (instead of error-reduction algorithm), former has been reported to have better convergence speed and  immunity to the stagnation problem \cite{fienup1982phase, bauschke2002phase}.

In conclusion, this letter demonstrated that a deep neural network trained for the image restoration task can also be used for solving the ill-posed phase retrieval problem (from a single diffracted intensity image). This transfer of knowledge was achieved here because the laws of diffraction physics were enforced on the output of the deep neural network. The method did not require any ground truth phase imaging data in any stage. The reconstruction quality was better as compared to the reconstruction obtained with the total-variation prior. Additionally, it was concluded that the joint use of handcrafted priors, such as the TV prior and deep neural networks can facilitate a faithful estimation in much lesser time, for a recurrent physics enforced deep learning framework.

\section*{Disclosures}
The author declares no conflicts of interest.

\bibliography{PhyZSN_OL}

\begin{thebibliography}{10}
\newcommand{\enquote}[1]{``#1''}

\bibitem{fienup1982phase}
J.~R. Fienup, {\protect\JournalTitle{Applied optics}} \textbf{21}, 2758 (1982).

\bibitem{rivenson2018phase}
Y.~Rivenson, Y.~Zhang, H.~G{\"u}nayd{\i}n, D.~Teng, and A.~Ozcan,
  {\protect\JournalTitle{Light: Science \& Applications}} \textbf{7}, 17141
  (2018).

\bibitem{gerchberg1972practical}
R.~W. Gerchberg, {\protect\JournalTitle{Optik}} \textbf{35}, 237 (1972).

\bibitem{latychevskaia2007solution}
T.~Latychevskaia and H.-W. Fink, {\protect\JournalTitle{Physical review
  letters}} \textbf{98}, 233901 (2007).

\bibitem{mudanyali2010compact}
O.~Mudanyali, D.~Tseng, C.~Oh, S.~O. Isikman, I.~Sencan, W.~Bishara,
  C.~Oztoprak, S.~Seo, B.~Khademhosseini, and A.~Ozcan,
  {\protect\JournalTitle{Lab on a Chip}} \textbf{10}, 1417 (2010).

\bibitem{chambolle2004algorithm}
A.~Chambolle, {\protect\JournalTitle{Journal of Mathematical imaging and
  vision}} \textbf{20}, 89 (2004).

\bibitem{bioucas2007new}
J.~M. Bioucas-Dias and M.~A. Figueiredo, {\protect\JournalTitle{IEEE
  Transactions on Image processing}} \textbf{16}, 2992 (2007).

\bibitem{katkovnik2012high}
V.~Katkovnik and J.~Astola, {\protect\JournalTitle{JOSA A}} \textbf{29}, 44
  (2012).

\bibitem{jolivet2018regularized}
F.~Jolivet, F.~Momey, L.~Denis, L.~M{\'e}{\`e}s, N.~Faure, N.~Grosjean,
  F.~Pinston, J.-L. Mari{\'e}, and C.~Fournier, {\protect\JournalTitle{Optics
  express}} \textbf{26}, 8923 (2018).

\bibitem{rivenson2019deep}
Y.~Rivenson, Y.~Wu, and A.~Ozcan, {\protect\JournalTitle{Light: Science \&
  Applications}} \textbf{8}, 1 (2019).

\bibitem{wu2018extended}
Y.~Wu, Y.~Rivenson, Y.~Zhang, Z.~Wei, H.~G{\"u}naydin, X.~Lin, and A.~Ozcan,
  {\protect\JournalTitle{Optica}} \textbf{5}, 704 (2018).

\bibitem{misell1973examination}
D.~Misell, {\protect\JournalTitle{Journal of Physics D: Applied Physics}}
  \textbf{6}, 2200 (1973).

\bibitem{paganin2004quantitative}
D.~Paganin, A.~Barty, P.~McMahon, and K.~A. Nugent,
  {\protect\JournalTitle{Journal of microscopy}} \textbf{214}, 51 (2004).

\bibitem{wang2019net}
K.~Wang, J.~Dou, Q.~Kemao, J.~Di, and J.~Zhao, {\protect\JournalTitle{Optics
  letters}} \textbf{44}, 4765 (2019).

\bibitem{ulyanov2018deep}
D.~Ulyanov, A.~Vedaldi, and V.~Lempitsky, \enquote{Deep image prior,} in
  \emph{Proceedings of the IEEE conference on computer vision and pattern
  recognition,}  (2018), pp. 9446--9454.

\bibitem{wang2020phase}
F.~Wang, Y.~Bian, H.~Wang, M.~Lyu, G.~Pedrini, W.~Osten, G.~Barbastathis, and
  G.~Situ, {\protect\JournalTitle{Light: Science \& Applications}} \textbf{9},
  1 (2020).

\bibitem{barbastathis2019use}
G.~Barbastathis, A.~Ozcan, and G.~Situ, {\protect\JournalTitle{Optica}}
  \textbf{6}, 921 (2019).

\bibitem{karniadakis2021physics}
G.~E. Karniadakis, I.~G. Kevrekidis, L.~Lu, P.~Perdikaris, S.~Wang, and
  L.~Yang, {\protect\JournalTitle{Nature Reviews Physics}} pp. 1--19 (2021).

\bibitem{raissi2019physics}
M.~Raissi, P.~Perdikaris, and G.~E. Karniadakis, {\protect\JournalTitle{Journal
  of Computational Physics}} \textbf{378}, 686 (2019).

\bibitem{heaven2019deep}
D.~Heaven, {\protect\JournalTitle{Nature}} \textbf{574}, 163 (2019).

\bibitem{psichogios1992hybrid}
D.~C. Psichogios and L.~H. Ungar, {\protect\JournalTitle{AIChE Journal}}
  \textbf{38}, 1499 (1992).

\bibitem{chan2021rapid}
H.~Chan, Y.~S. Nashed, S.~Kandel, S.~O. Hruszkewycz, S.~K. Sankaranarayanan,
  R.~J. Harder, and M.~J. Cherukara, {\protect\JournalTitle{Applied Physics
  Reviews}} \textbf{8}, 021407 (2021).

\bibitem{goodman2005introduction}
J.~W. Goodman, \emph{Introduction to Fourier optics} (Roberts and Company
  Publishers, 2005).

\bibitem{born2013principles}
M.~Born and E.~Wolf, \emph{Principles of optics: electromagnetic theory of
  propagation, interference and diffraction of light} (Elsevier, 2013).

\bibitem{wang2019survey}
W.~Wang, V.~W. Zheng, H.~Yu, and C.~Miao, {\protect\JournalTitle{ACM
  Transactions on Intelligent Systems and Technology (TIST)}} \textbf{10}, 1
  (2019).

\bibitem{xian2018zero}
Y.~Xian, C.~H. Lampert, B.~Schiele, and Z.~Akata, {\protect\JournalTitle{IEEE
  transactions on pattern analysis and machine intelligence}} \textbf{41}, 2251
  (2018).

\bibitem{zhang2020plug}
K.~Zhang, Y.~Li, W.~Zuo, L.~Zhang, L.~Van~Gool, and R.~Timofte,
  {\protect\JournalTitle{arXiv preprint arXiv:2008.13751}}  (2020).

\bibitem{zhang2017learning}
K.~Zhang, W.~Zuo, S.~Gu, and L.~Zhang, \enquote{Learning deep cnn denoiser
  prior for image restoration,} in \emph{Proceedings of the IEEE conference on
  computer vision and pattern recognition,}  (2017), pp. 3929--3938.

\bibitem{bauschke2002phase}
H.~H. Bauschke, P.~L. Combettes, and D.~R. Luke, {\protect\JournalTitle{JOSA
  A}} \textbf{19}, 1334 (2002).

\bibitem{ozcan2016lensless}
A.~Ozcan and E.~McLeod, {\protect\JournalTitle{Annual review of biomedical
  engineering}} \textbf{18}, 77 (2016).

\bibitem{burri2019}
O.~Burri and R.~Guiet,
  {\protect\JournalTitle{http://doi.org/10.5281/zenodo.3232478}}  (2019).

\end{thebibliography}


\bibliographyfullrefs{PhyZSN_OL}



\ifthenelse{\equal{\journalref}{aop}}{%
\section*{Author Biographies}
\begingroup
\setlength\intextsep{0pt}
\begin{minipage}[t][6.3cm][t]{1.0\textwidth} 
  \begin{wrapfigure}{L}{0.25\textwidth}
    \includegraphics[width=0.25\textwidth]{john_smith.eps}
  \end{wrapfigure}
  \noindent
  {\bfseries John Smith} received his BSc (Mathematics) in 2000 from The University of Maryland. His research interests include lasers and optics.
\end{minipage}
\begin{minipage}{1.0\textwidth}
  \begin{wrapfigure}{L}{0.25\textwidth}
    \includegraphics[width=0.25\textwidth]{alice_smith.eps}
  \end{wrapfigure}
  \noindent
  {\bfseries Alice Smith} also received her BSc (Mathematics) in 2000 from The University of Maryland. Her research interests also include lasers and optics.
\end{minipage}
\endgroup
}{}

\end{document}